\documentclass[twocolumn,showpacs,preprintnumbers,amsmath,amssymb]{revtex4}
\usepackage{graphicx}
\usepackage{dcolumn}
\usepackage{bm}
\begin{document}
\title{Fragmentation paths in dynamical models}
\author{M. Colonna$^{a}$, A. Ono$^{b}$, J. Rizzo$^{a}$} 
\affiliation{
        $^a$LNS-INFN, I-95123, Catania, Italy\\
	$^b$Department of Physics, Tohoku University, Sendai 980-8578, Japan}
\begin{abstract}
We undertake a quantitative comparison of multi-fragmentation reactions, as modeled
by two different approaches: the Antisymmetrized Molecular Dynamics (AMD) and the
momentum-dependent stochastic mean-field (SMF) model.
Fragment observables  and pre-equilibrium (nucleon and light cluster) emission
are analyzed, in connection to the 
underlying compression-expansion dynamics in each model. 
Considering reactions between neutron-rich systems, observables related 
to the isotopic properties of emitted particles and fragments are also discussed, as a function
of the parametrization employed for the isovector part of the nuclear interaction.  
We find that the reaction 
path, particularly the mechanism of fragmentation, is different in the 
two models and reflects on some properties of the reaction 
products, including their 
isospin content. This should be taken into account in the study of  
the density dependence of the symmetry energy from such collisions.
\end{abstract}
\maketitle

\section{Introduction}

Due to the analogies between the nuclear forces and the Van-der-Waals interaction,  
the nuclear matter equation of state (EOS) foresees the possible occurrence of phase transitions 
from the liquid to the vapour phases \cite{report_Borderie,rep}. Already in the 80's, these phenomena were linked to the
experimentally observed multifragmentation mechanism, i.e.\ the sudden break up of an
excited nuclear system into many pieces \cite{Bow,Bert}. However, in the search for signatures of a phase transition,  
one has to consider that nuclei are finite systems. 
New features may appear 
in addition to
standard thermodynamics. Moreover, the nuclear phase diagram
is explored with the help of nuclear reactions, where out-of-equilibrium effects may also be present. 
These considerations have lead, from one side, to significant developments of the thermodynamics
of finite systems and to new definitions of the phase transition, following the
concepts of statistical physics \cite{Chomaz,Chomaz_new}. On the other hand, many efforts have been devoted to the study
of the reaction dynamics and to the characterization of the fragmentation mechanism, 
also in connection with the possible appearance of the 
signals expected for a phase transition at thermodynamical equilibrium 
\cite{Moretto,Dagostino,EPJA,Xu00,Ger04,Ono_Tak,Alexandru}.

It is generally believed that, in a heavy ion collision at Fermi energies, due to the initial
compression and/or thermal excitation, the composite nuclear system may expand and reach density and
temperature values inside the co-existence region of the nuclear matter phase diagram \cite{rep,EPJA,Ono}.  
In violent collisions  
multifragment configurations are characterized by a large degree of chaoticity 
and a huge amount of the available phase space is populated.
Finally several features 
observed in the exit channel can be related
to the thermodynamical (equilibrium) properties of phase transitions in finite systems, independently of the 
specific mechanism that has driven the whole process \cite{Alexandru,Ono_Tak}. 
However, there remain some aspects that reflect the nature of the reaction dynamics and of the fragmentation
path, mostly related to out-of-equilibrium effects. Indeed the interplay between the compression-expansion 
dynamics and the onset (and nature) of clusterization affects significantly
the fragment kinematical properties
and the appearance of collective flow effects. A thorough study of this dynamics also allows one to access information
on the nuclear matter compressibility.   Moreover, the specific features of the fragmenting source, such as its mass,
charge and excitation energy, are also strongly depending on the pre-equilibrium dynamics and on the time instant
where clusterization sets in. 

Hence, a quantitative understanding of multifragmentation data requires a careful investigation 
of the whole dynamical path. Due to the complexity of the nuclear many-body problem, 
two main lines of approximations have been followed so far. On one side, 
the class of molecular dynamics (MD) approaches employ, to represent a many-body state,  
a product of single-particle states, with or without 
antisymmetrization, where only the mean positions and momenta are time 
dependent \cite{md,feld,ONOab,Pap,FFMD,ImQMD}. 
In almost all these approaches, the width is fixed and is the same for all wave packets. 
The use of localized wave packets
induces many-body correlations (analogous to those in classical dynamics)
in the particle propagation in the nuclear field, as well as in  
hard two body scattering, that is treated stochastically. 
On the other side, mean-field approaches (and stochastic extensions) follow the
time evolution of the one-body distribution function  (the semi-classical analog
of the Wigner transform of the one-body density matrix), according to
approximate equations where higher order correlations are neglected 
(mean-field approximation), apart from the correlations  
introduced by the residual two-body collisions. 
In the stochastic extension of the model, 
a fluctuation source term
is added to the average collision integral, to account for the stochastic nature of 
two-body scattering \cite{Ayik,Randrup,rep,BL_new}.  
    
A comparison of the predictions given by models of the two classes, concerning
multifragmentation scenarios at Fermi energies (30-50 MeV/nucleon), has been recently
undertaken \cite{comp_1}. 
The models considered are: the stochastic mean-field (SMF) model
including momentum dependence \cite{Joseph_model} and the 
antisymmetrized molecular dynamics (AMD) model \cite{Ono}.
It was observed that, while in the SMF case 
fragment  emission is more likely connected to the spinodal decomposition mechanism,
i.e.\ to mean-field instabilities, in the AMD approach many-body correlations have
a stronger impact on the fragmentation dynamics, leading to the earlier development  
of density and momentum fluctuations. 

In the present paper we will discuss quantitatively the impact of the different approximations
employed in the two models  on several particle and fragment observables of experimental interest.
This also allows one to establish
up to which extent some of the observed features may be considered as more robust or general, i.e.\
not much depending  on the details of the fragmentation path and of the models
under consideration. By studying reactions with neutron-rich systems, we also investigate
isospin observables, in connection to
the density dependence of the symmetry energy 
and the underlying reaction dynamics.


\section{Description of the models}

\subsection{Basic framework}

In both SMF and AMD approaches, the time evolution of the system is
described in terms of the one-body distribution function $f$ (or a
Slater determinant), as ruled by the nuclear mean-field (plus Coulomb
interaction for protons) and the residual interaction, i.e.\ hard
two-body scattering.  The equation can be represented in the form of
so-called Boltzmann-Langevin equation \cite{Ayik,Randrup}
\begin{equation}
\frac{\partial f}{\partial t}
= \{H[f],f\} + I[f] 
+ \delta I[f],
\label{BL}
\end{equation}
where the coordinates of spin, isospin and phase space are not shown
for brevity.  $H[f]$ is the self-consistent one-body Hamiltonian,
$I[f]$ is the average two-body collision integral, and $\delta I[f]$
stands for the stochastic source term \cite{Ayik,Randrup}.

The effect of the stochastic term $\delta I[f]$ is considered for the
value of the distribution function $f_\alpha$ in each phase space cell
$\alpha$.  For example, one may decompose the phase space into square
cells of the volume $(2\pi\hbar)^3$.  In the case of local
equilibrium, the source term $\delta I[f]$ induces the fluctuation of
$f_\alpha$, the variance of which is given by $\langle\Delta
f_\alpha^2\rangle=f_\alpha(1-f_\alpha)$.  The correlations between
different cells are usually assumed to be small, but minimal
correlations are introduced for the conservation laws.  In the
approximate treatment of SMF presented in \cite{Alfio,Salvo}, the
fluctuations are projected and implemented only onto the ordinary
space, agitating the spatial density profile from time to time during
the reaction after the local thermal equilibrium is reached.

It should be noted that the variance $\langle\Delta
f_\alpha^2\rangle=f_\alpha(1-f_\alpha)$ with the average value
$f_\alpha$ is achieved by choosing $f_\alpha+\Delta f_\alpha=1$ and 0
with the probabilities $f_\alpha$ and $1-f_\alpha$, respectively.
Namely the fluctuation will choose $A$ fully occupied phase-space
cells, with $A$ being the number of nucleons in the system, while
the remaining part of phase space becomes empty.  AMD represents this
situation by using Gaussian wave packets $\propto
e^{-2\nu(\mathbf{r}-\mathbf{R}_i)^2-(\mathbf{p}-\mathbf{P}_i)^2/2\hbar^2\nu}$
($i=1,\ldots,A$).  Each wave packet has the minimum uncertainty
$\Delta x\Delta p=\frac{1}{2}\hbar$ and therefore can be regarded as a
phase space cell.  The width parameter $\nu=(2.5\ \textrm{fm})^{-2}$
defines the shape of the cell in phase space.  

In AMD, the right hand side of Eq.\ (\ref{BL}) is usually decomposed
in a different way for the convenience in writing a stochastic
equation for the wave packet centroids.  The mean field effect is
decomposed to the motion of the centroids and the change of the shape
of the one-body distributions:
$\{H,f\}=\{H,f\}_{\text{cent}}+\{H,f\}_{\text{shape}}$.  The first
term $\{H,f\}_{\text{cent}}$ is calculated by employing the fully
antisymmetrized wave function \cite{ONOab}, while
$\{H,f\}_{\text{shape}}$ is approximately calculated by using a test
particle method \cite{Onoj,ONO-ppnp}.  The stochastic source term is
also separated as $\delta I=\delta I_{\text{coll}}+\delta
I_{\text{split}}$.  The part $\delta I_{\text{coll}}$ is the
stochastic effect included in the 
stochastic collisions  ($I+\delta I_{\text{coll}}$),
that move the centroids of the wave packets by choosing the scattering
angle randomly. When  a nucleon collides with
another, $\delta I_{\text{split}}$ randomly selects a Gaussian wave
packet from the single-particle state, that has been changing its
shape according to $\{H,f\}_{\text{shape}}$, by splitting it into
possibilities of Gaussian wave packets (quantum branching or
decoherence) \cite{Onoj,ONO-ppnp}.



Thus the main difference between the two models lies in the
implementation of the stochastic term, i.e., whether the fluctuation
in the momentum space is integrated out (in SMF) or not (in AMD).  
Moreover, 
in AMD, when a collision happens, two entire nucleons
are moved to new wave packets in phase space,
while it is not necessarily the case in SMF.  Therefore the
fluctuation is expected to have stronger impact on the collision
dynamics in AMD than in SMF.

\subsection{Effective interaction}

The isospin and momentum dependent effective interaction employed in
the SMF model is derived via an asymmetric extension of the GBD force
\cite{GalePRC41,GrecoPRC59}, leading to the BGBD potential
\cite{Isospin01,Bombiso,Joseph_model}.  In the AMD model, we employ
the Gogny interaction which are composed of finite-range two-body
terms and a zero-range density-dependent term.

\begin{figure}
\includegraphics[width=6.cm]{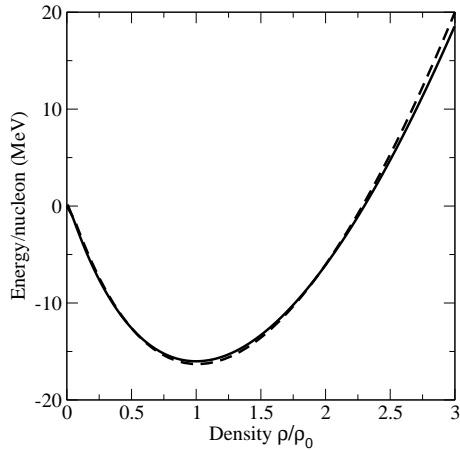}
\caption{Equation of state (EOS) of symmetric nuclear matter,
  corresponding to the Gogny (full) and BGBD (dashed) interactions.}
\label{eos}
\vskip 0.5cm
\end{figure}

\begin{figure}
\includegraphics[width=6.cm]{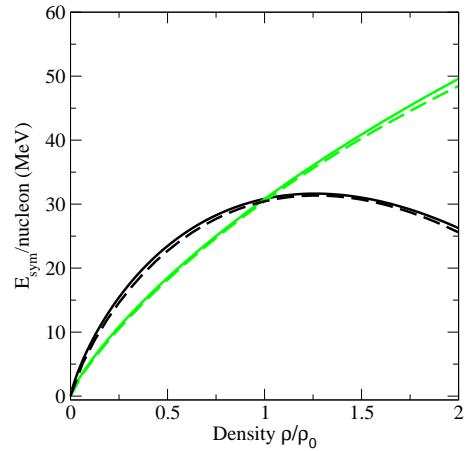}
\caption{ Parametrizations of the density behaviour of the symmetry
  energy adopted in the calculations: Asy-soft (black lines),
  Asy-stiff (grey lines). Full lines correspond to the Gogny forces,
  while dashed lines are for the BGBD forces.}
\label{esy}
\vskip 1.cm
\end{figure}

The corresponding nuclear matter EOS can be written as
$(E/A)(\rho,\beta) = (E/A)(\rho)+ E_{\text{sym}} (\rho) \beta^2 +
O(\beta^4) +..,$ where the variable $\beta=(\rho_n-\rho_p)/\rho$
defines the isospin content, being $\rho_\tau$ ($\tau$ = n, p) the
neutron or proton density and $\rho$ the total density.  We use a soft
equation of state for symmetric nuclear matter (compressibility
modulus $K_{\text{NM}}(\rho_0)=215$ MeV for BGBD and
$K_{\text{NM}}=228$ MeV for the Gogny force).  In Fig.~\ref{eos} we
report the EOS, as a function of the density $\rho$, for symmetric
nuclear matter, $\beta = 0$, at zero temperature.  We can easily
adjust the parameters of the interaction in order to change the
density dependence of symmetry energy ($E_{\text{sym}}$) without
changing the EOS of symmetric nuclear matter.  The density behaviour
of the symmetry energy, $E_{\text{sym}}(\rho)$, is shown in
Fig.~\ref{esy}.  We remind that this quantity gets a kinetic
contribution directly from basic Pauli correlations and a potential
part from the highly controversial isospin dependence of the effective
interactions \cite{BaranPR410,EPJA_betty}.  For the BGBD force, we
adopt two different parametrizations of the symmetry energy:
``asy-soft'', that gives a flat behaviour around normal density,
followed by a decreasing trend at large density, and ``asy-stiff'',
where the symmetry energy exhibits an almost linear increase with
density \cite{Baran_force}.  The Gogny (D1) force \cite{GOGNY} has a quite similar
$E_{\text{sym}}(\rho)$ to the ``asy-soft'' parametrization of BGBD,
while another parametrization \cite{ONOk} of the Gogny force has
$E_{\text{sym}}(\rho)$ similar to the ``asy-stiff'' version of BGBD.
Hence, in the following we will use the labels ``asy-soft'' and
``asy-stiff'' for the Gogny forces as well.


Within the considered form of the nuclear interaction, the isovector
terms depend explicitely on the nucleon momentum $k = p/\hbar$,
leading to the splitting of neutron and proton effective masses.  The
mean-field potential felt by neutrons and protons in asymmetric
nuclear matter ($\beta = 0.2$) is presented in Fig.~3, as a function
of the momentum $k$, for three density values and for the adopted
interactions (asy-soft and asy-stiff).  As an effect of the momentum
dependence, the difference between neutron and proton potentials
decreases with $k$, becoming negative at high momenta, especially in
the asy-soft case.  This corresponds to the proton effective mass
being smaller than the neutron one in neutron-rich matter.  Moreover,
one can see that the difference between neutron and proton potentials
is larger, at low density, in the asy-soft case, corresponding to the
larger value of the symmetry energy (see Fig.~\ref{esy}), while the
opposite holds above normal density.  The results of the BGBD and
Gogny forces look very close to each other below and around the Fermi
momentum, for both neutrons and protons.  However, due to the
different shape of the momentum dependence in the BGBD interaction,
with respect to the Gogny force, at high momenta ($k \geq
3~\textrm{fm}^{-1}$) potential are less attractive in the BGBD
case. This discrepancy should not affect our comparison since we are
interested in reactions at Fermi energies.

\begin{figure}[t]
\includegraphics[width=8.cm]{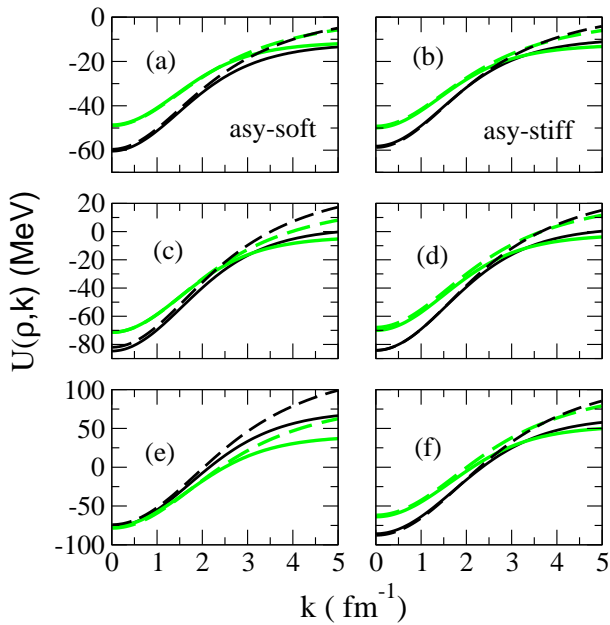}
\caption{Mean-field potential, for neutrons (grey lines) and protons
  (black lines) in asymmetric matter ($\beta = 0.2$) at density $\rho
  = 0.085~\textrm{fm}^{-3}$ (panels (a) and (b)), $\rho =
  0.17~\textrm{fm}^{-3}$ (panels (c) and (d)), $\rho =
  0.34~\textrm{fm}^{-3}$ (panels (e) and (f)), as a function of the
  momentum $k$.  Full lines correspond to the Gogny forces while
  dashed lines are for the BGBD forces.  Left panels: asy-soft
  interaction. Right panels: asy-stiff.}
\label{udik}
\end{figure}

\subsection{Two-nucleon collision cross sections}

As the two-nucleon collision cross sections ($\sigma_{pp}=\sigma_{nn}$
and $\sigma_{pn}$), we use the energy- and angle-dependent values in
the free space with the maximum cut-off of 150 mb in both SMF and AMD
calculations.  We have confirmed that the degree of stopping reached
in the reaction is similar for both calculations \cite{comp_1}.
However, it should be noticed that these cross sections are larger
than those adopted by the AMD calculation in Ref.\ \cite{Onoj}.

\begin{figure*}[t]
\includegraphics[width=18.cm]{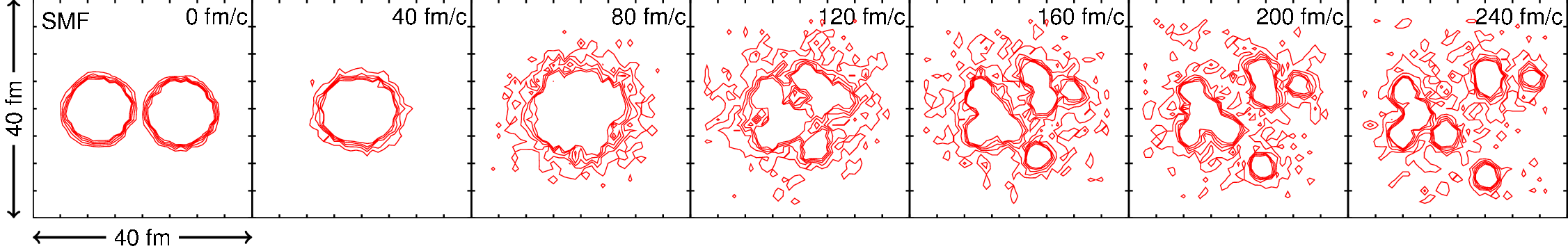}
\vskip 0.5cm
\includegraphics[width=18.cm]{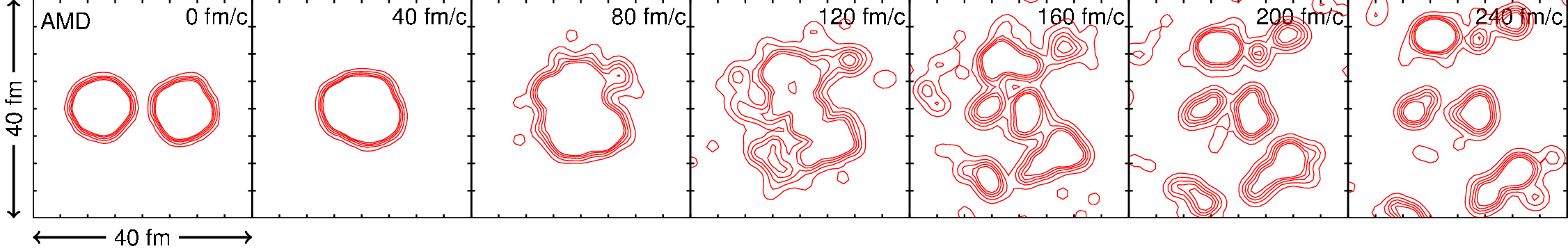}
\caption{Contour plots of the density projected on the reaction plane
  calculated with SMF (top) and AMD (bottom) for the central reaction $^{112}$Sn +
  $^{112}$Sn at 50 MeV/nucleon, at several times (fm/$c$). 
The lines are drawn at projected
   densities beginning at 0.07 $fm^{-2}$ and increasing by 0.1 $fm^{-2}$.
The size of each box is 40 fm.
}
\label{contour_BGBD}
\end{figure*}

\subsection{Fragment identification}

In the SMF model, the reaction products are reconstructed by applying
a coalescence procedure to the one-body density $\rho({\bf r})$, i.e.\
connecting neighboring cells with density $\rho \geq
\frac{1}{6}\rho_0$ (``liquid'' phase).  In this way one can also
identify a ``gas'' phase ($\rho < \frac{1}{6}\rho_0$), associated with
particles that leave rapidly the system (pre-equilibrium emission)
and/or are evaporated.  Once fragments are identified, from the
knowledge of the one-body distribution function, it is possible to
calculate their mass, charge and kinematical properties.  This
provides reasonable results for the description of the ground state of
medium-heavy nuclei and for excited primary fragment properties.
Fragment excitation energies are calculated by subtracting the Fermi
motion, evaluated in the local density approximation, from the
fragment kinetic energy (taken in the fragment reference frame)
\cite{rep,Alfio}.

In the AMD model, for the state at a given time of the reaction,
fragments are recognized by connecting two wave packets if the spatial
distance between their centroids is less than $d_{\text{cl}}$.  When
fragments are recognized at a late time, such as $t=200$ fm/$c$, we
take $d_{\text{cl}}=5$ fm, though the result does not depend on the
choice except for extreme choices.  When fragments are recognized at
early times in order to characterize the dynamical evolution of the
reaction, different choices of $d_{\text{cl}}$ define different
observables.  We take a relatively small value of $d_{\text{cl}}=3$ fm
in Figs.\ \ref{pre_eq_124}, \ref{pre_eq_112}, \ref{nzgas} and
\ref{z_a_liq}, which allows an earlier recognition of particle
emissions.

Alternatively, we can employ the same fragment recognition method that
is applied to the SMF calculation by applying a coalescence procedure
to the density distribution $\rho(\mathbf{r})$ calculated for the AMD
wave function.  We have confirmed that the result well agrees with the
usual method with $d_{\text{cl}}=5$ fm for fragment charge
distributions (i.e., for the separation of fragments), while the choice
of $d_{\text{cl}}=3$ fm better agrees with the coalescence procedure
for the light particle emissions (i.e., for the separation of
``liquid'' and ``gas'') at early times.

\section{Results}

We discuss some features of the fragmentation path followed  
in violent collisions at Fermi energies, as predicted by the SMF and the AMD models.
To investigate isospin effects and the sensitivity of the results to the parametrization
adopted for the symmetry energy, we will simulate central collisions of 
a neutron-poor and a neutron-rich system: $^{112}$Sn +
$^{112}$Sn and $^{124}$Sn + $^{124}$Sn, at 50 MeV/nucleon.
Within our study of fragmentation reactions, we expect to test the low-density behaviour
of the symmetry energy. 
As we will see in the following,
the isospin degree of freedom can also be used as a good tracer of the reaction dynamics. 
  An ensemble of about 100 trajectories 
has been collected for both model calculations.
The impact parameter has been set equal to 0.5 fm and the initial distance
between the two nuclei (at the time $t = 0$) is 15 fm.
Density contour plots in the reaction plane, as obtained in
the two models for one event of the  $^{112}$Sn +
$^{112}$Sn reaction, are shown in Fig.\ \ref{contour_BGBD} 
at several time steps.  As one can see from the figure,
for this kind of reactions, both models predict that the system is initially
compressed. Then expansion follows and multi-fragment breakup is observed. 
One can also observe 
that the expansion is faster in AMD and that 
the amount of emitted nucleons (the ``gas'' phase)
is larger in SMF.   

\subsection{Compression-expansion dynamics}

\begin{figure}
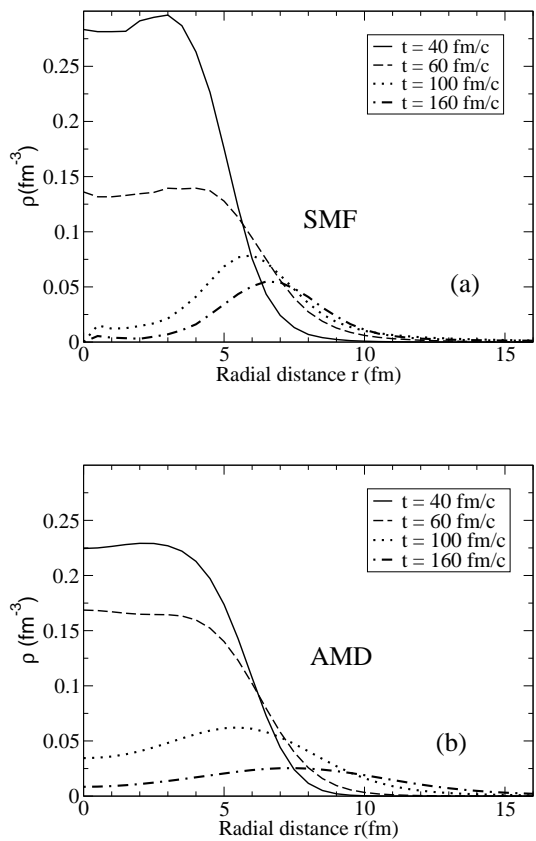

\vskip 1.cm
\includegraphics[width=7.cm]{fig5a_com.eps}
\vskip 1.cm
\includegraphics[width=7.cm]{fig5b_com.eps}
\caption{Density profiles, at several times, as obtained in the 
SMF (a) and AMD (b) case, for the reaction  $^{112}\mathrm{Sn} + {}^{112}\mathrm{Sn}$.}
\label{dens_SMF}
\vskip 0.5cm
\end{figure} 

 
In this Section we summarize the main findings of Ref.~\cite{comp_1}, concerning
the trajectory followed by the system in the early stage of the dynamics, until 
fragmentation is observed. 
The reaction path can be characterized with the help of  
one-body observables, such as  the radial density profile and
the radial collective momentum. 
The radial density at a given distance $r$ is obtained by averaging the local density $\rho({\bf r})$
over the surface of a sphere of radius $r$. 
The radial collective momentum is the projection of the collective momentum at the position ${\bf r}$
along the radial direction, averaged over the surface of the sphere of radius $r$.
These quantities are further averaged over the ensemble events.  

In the SMF calculations (see Fig.~5(a))
the behaviour of the radial density profile indicates that, after an initial compression (t = 40 fm/c), the
system expands and finally it gets rather dilute, due to the occurrence of 
a monopole expansion, generated by the initial compression.   
The matter appears mostly 
concentrated within a given interval of the radial distance (see for instance the results
at t = 100 fm/c), indicating the formation
of bubble-like configurations (see also  Fig.\ \ref{contour_BGBD}), where fragments will appear, and 
the central region of the system is rapidly depleted.

In AMD calculations, the density profile in Fig.~5(b) shows
the time evolution of the compression and expansion which is
qualitatively similar to the SMF case.  However, we notice that AMD
shows broader average density distribution than SMF as the system expands,
pointing to
a faster expansion in the AMD, as indicated qualitatively by Fig.
\ref{contour_BGBD}. This is confirmed also by
the behavior  of the collective momentum.
An shown in Ref.~\cite{comp_1},
in the SMF case, after the development of an almost self-similar radial
flow, the collective momentum decreases again at the surface of the system,
indicating the occurrence of a counter-streaming flow, from the
surface towards the interior, trying to recompact the system 
\cite{Batko}. Correspondingly a bump is obtained in the radial density
distribution (see Fig.~5(a) at t = 100 fm/c).
On the other hand,
the collective
momentum at the surface of the expanding system
keeps almost unchanged in the AMD calculations \cite{comp_1}.
The absence of deceleration effects may 
indicate that the system ceases to behave as homogeneous while
it expands,  
corresponding to 
the situation, of rather low average density,  in which fragments have already appeared and 
are distributed widely in space.
This scenario is supported by 
the analysis of the density fluctuation variance, that
reveals an earlier growth of density fluctuations (leading to an earlier appearance
of clusters) in the AMD case \cite{comp_1}.  
This can be interpreted
     as the reason of the faster expansion in AMD as observed in Fig.\ \ref{contour_BGBD} 
     at later times.
We notice that  early fragment formation is observed 
also in other N-body treatments, belonging to the class of molecular
dynamics models, as shown by the quasi-classical calculations
performed in Ref.~\cite{Dorso}.

In conclusion, while
in the SMF case the system spends more time, as a nearly homogeneous
source, at low density and fragments are formed most likely through
the spinodal mechanism \cite{rep}, in the AMD case clustering effects
may be present at an earlier stage.   
This influences significantly also the amount of
particles emitted prior to fragment formation (pre-equilibrim emission),
as discussed in the following. 


\subsection{Time evolution of nucleon and cluster emission}

\begin{figure}
\includegraphics[width=8.cm]{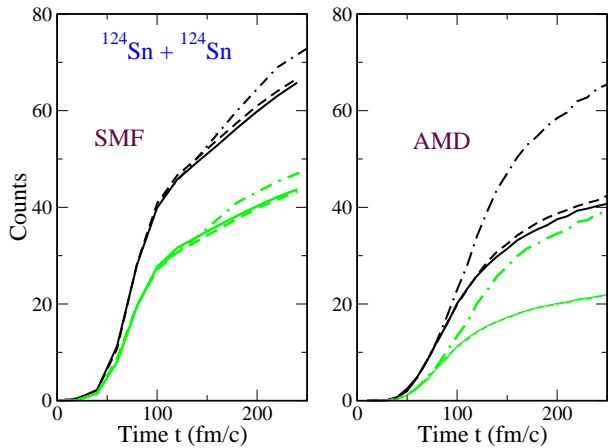}
\caption{
Time evolution of the total
  number of neutrons (black lines) and 
protons (grey lines) 
contained in emitted particles with $1\leq A\leq 4$, in the case
of the reaction $^{124}\mathrm{Sn} + {}^{124}\mathrm{Sn}$. Dashed line: asy-soft parametrization. 
Full line: asy-stiff parametrization.
The dot-dashed lines represent the time evolution of the total number of neutrons and
protons contained in emitted particles with $1\leq A\leq 15$, for the asy-soft parametrization.
Left panel: SMF results. Right panel: AMD results. }  
\label{pre_eq_124}
\end{figure}

\begin{figure}
\vskip 1.0cm
\includegraphics[width=8.cm]{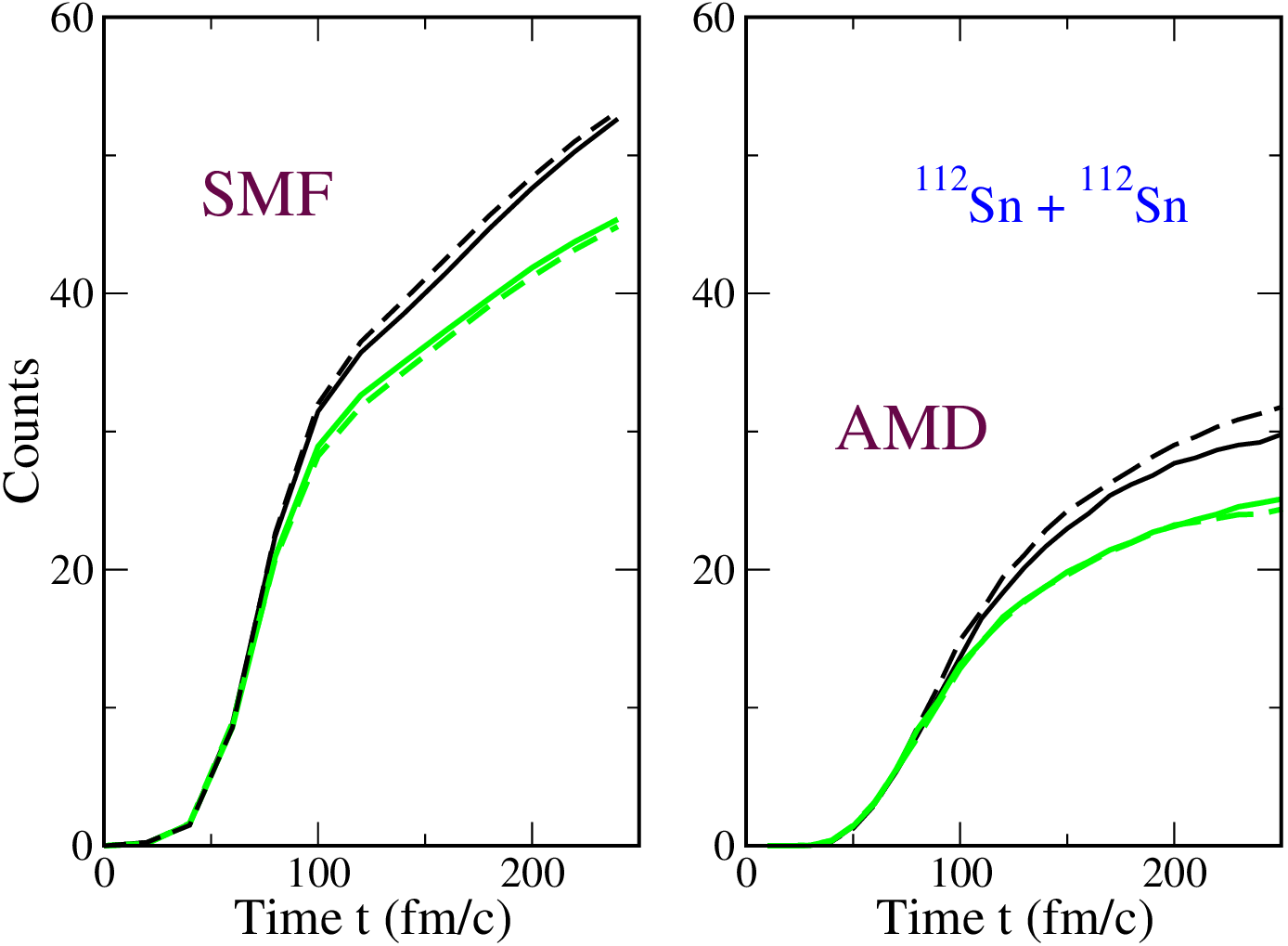}
\caption{The same as in Fig.~\ref{pre_eq_124}, for the system  $^{112}\mathrm{Sn} + {}^{112}\mathrm{Sn}$.}
\label{pre_eq_112}
\end{figure}

During the first stage of the reaction, hard two-body scattering plays an essential role 
and pre-equilibrium emission is observed, i.e.\ nucleons and light particles are promptly emitted
from the system. This stage  influences significantly the
following evolution of the collision. In fact, the amount of particles and energy
removed from the system affects the properties of the composite source that eventually
breaks up into pieces. Hence, when discussing multifragmentation mechanisms, a detailed
analysis of this early emission is in order.
Figs.~\ref{pre_eq_124} and \ref{pre_eq_112} show the time evolution of the total number of neutrons and protons contained in
emitted nucleons and light particles, with  mass number $A\leq 4$, as obtained 
in the two models, with the two asy-EOS considered, and for the two reactions. 
It should be noticed that this ensemble is mainly composed  of unbound nucleons  
in the SMF case. 
As shown in these figures, these particles leave the system mostly in the
time interval between $\approx 70$ fm/$c$ and $\approx 120$ fm/$c$. At later times the emission rate
is reduced (see the change of slope in the lines of Figs.~\ref{pre_eq_124} 
and \ref{pre_eq_112}.
As already pointed out in Ref.~\cite{comp_1}, a striking difference between the two models
concerns the amount of particles emitted, that is larger in the SMF case. 
However, the average kinetic energy of this emission is similar,  
being 20.72 MeV/nucleon in SMF and  21.95 MeV/nucleon in AMD. 
The difference observed  
in the two models could be connected to the fact that 
clustering effects and many-body correlations 
are more efficient in AMD, due to the nucleon localization,
reducing the amount of mass that goes
into free nucleons and very light clusters. 
This effect is connected also to the different compression-expansion
dynamics in the two models, as discussed in the previous section. 

On the other hand, 
light IMF's, with mass number $5 \leq A\leq 15$,  
are more abundant in AMD.  This is observed in Fig.~\ref{pre_eq_124}, where the total amount
of neutrons and protons contained in emitted particles with $A \leq 15$ 
is also displayed as a function of time, in the case of the asy-soft 
interaction (see the dot-dashed lines). 
In the AMD case, the emission of light IMF's starts already at around 70 fm/$c   $
and, at the time t = 250 fm/$c$, it represents a noticeable   fraction
of the particles emitted in the considered mass range ($A \leq 15$),
see the difference between dot-dashed and dashed lines. 
Hence fragments are formed on shorter time scales in AMD, on about equal footing as light-particle pre-equilibrium emission,
while in SMF this light IMF emission sets in at later times and is much reduced.
However, one can see that 
the total amount of emitted nucleons belonging to particles 
with  $A \leq 15$ (including free nucleons) 
is close in the two models (compare the dot-dashed lines at the final time in the left and right
panels of Fig.~\ref{pre_eq_124}).  Hence one expects to see a similar production of
IMF's with charge  $A > 15$, as it will be discussed in the following.

As general features, 
when comparing the two reactions (Fig.~\ref{pre_eq_124} 
and Fig.~\ref{pre_eq_112}), it is seen that neutron (proton) emission is more
abundant in the neutron-rich (poor) systems.   Moreover, 
Figs.~\ref{pre_eq_124} and \ref{pre_eq_112}  also show that a larger (smaller) number of neutrons (protons)
is emitted in the asy-soft case, as compared with the asy-stiff case, 
corresponding to a larger repulsion of the symmetry potential for the soft parametrization.   
This can be taken as an indication of the fact that pre-equilibrium particles are mostly emitted  
from regions that are below normal density (i.e.\ after $t\approx 70$ fm/c, during the expansion phase), where the symmetry
energy is higher in the soft case (see Fig.~2).



\subsection{Fragment properties}
\begin{figure}
\vskip 1.0cm
\includegraphics[width=8.cm]{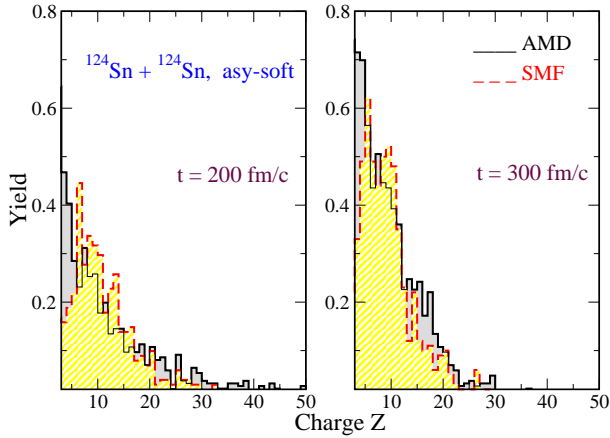}
\caption{Charge distribution as obtained in AMD (full histogram) 
and in SMF (dashed histogram), for the reaction $^{124}\mathrm{Sn} + {}^{124}\mathrm{Sn}$,
at $t = 200$ fm/$c$ (left) and $t = 300$ fm/$c$ (right).}
\label{char_distri}
\end{figure}

Now we move to discuss and compare the properties of primary fragments, as obtained in the two models. 
In Fig.~\ref{char_distri}  we present the charge distribution of primary IMF's ($Z>2$)
at two time instants: $t = 200$ and 300 fm/$c$.
\begin{figure}
\vskip 1.0cm
\includegraphics[width=6.cm]{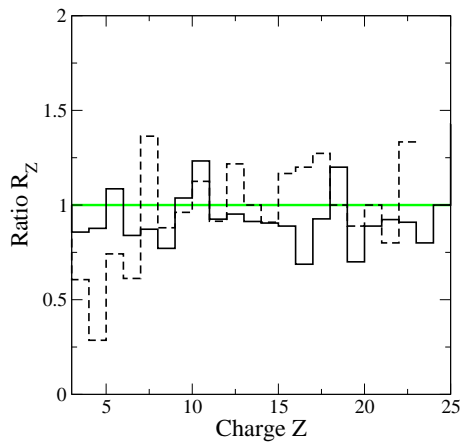}
\caption{
 Ratio $R_Z$ between 
the fragment yield observed in the angular domain $60 < \theta < 120$
(rescaled by a factor 2) 
and the total yield, 
as a function of the fragment charge, evaluated at $t=300$ fm/$c$
for the same system of  Fig.~\ref{char_distri}.
Notations are the same as in Fig.~\ref{char_distri}.} 
\label{char_perc}
\end{figure}
Some of the fragments identified by the clustering procedure at $t = 200$ fm/$c$ have exotic
(elongated) shapes and they break-up into pieces at later times. This effect is present
in both models and explains the difference observed between the charge distributions calculated 
at $t = 200$ and $t = 300$ fm/$c$.    
In the SMF case the maximum yield corresponds to fragments
having charge around $Z = 10$, as expected within a fragmentation scenario associated
with spinodal decomposition \cite{rep}.   
The fragmentation mechanism may be connected also to the small yield of the light IMF's that,
as also discussed in Section III.B,
is lower
in SMF, compared to AMD.  Hence
the total mass that belongs to the ``liquid'' phase (IMF's)
is reduced in SMF. This is the counterpart of the more abundant 
pre-equilibrium emission.
The production of sizeable clusters ($Z>6$) looks closer 
in the two calculations.
This is true 
especially at $t = 300$ fm/$c$, where the tail at large $Z$, that is observed at $t = 200$ fm/$c$ and is 
more pronounced in AMD calculations, disappears. However, the emission of fragments with charge
around 10 is slightly larger in SMF, while 
bigger fragments (with charge around 15-20) are more
abundant in AMD.  
It should be noticed that these differences may be smoothened by the 
secondary decay process, see the results of  Ref.\cite{frag} for SMF and
Ref.\cite{Onoj} for AMD.    

Fragments are nearly isotropically distributed in space, in both models. 
This is shown in Fig.~9, where we display the ratio, $R_Z$, 
between the fragment yield observed in the angular domain $60 < \theta < 120$
(rescaled by a factor 2) and the total yield, as a function of the fragment charge. 
Apart from fluctuations of statistical origin, we notice a lack of small fragments in the 
selected angular domain in the SMF case, pointing to a slightly lower degree of 
stopping, compared to AMD, as also discussed in Ref.~\cite{comp_1}.
We would like to stress that, especially in the case of SMF calculations, the degree
of stopping reached in the collision is crucial in determining the following reaction path. 
Indeed, if the reaction time becomes too short, spinodal instabilities would not have
enough time to develop \cite{rep}.
\begin{figure}
\vskip 1.0cm
\includegraphics[width=8.cm]{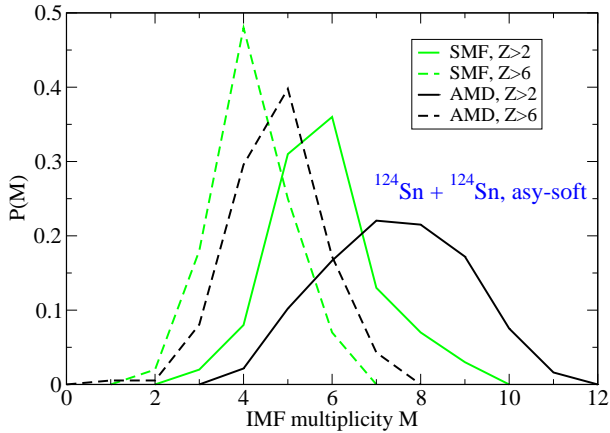}
\caption{IMF multiplicity distribution as obtained in the two models for the
same reaction of Fig.~\ref{char_distri}, at $t = 300$ fm/$c$. Full lines: IMF's
with charge $Z>2$ are considered. Dashed lines: IMF's with charge $Z>6$ are considered.}
\label{mult_distri}
\end{figure} 

The larger total mass of the ``liquid'' phase obtained in AMD is reflected also in the
IMF multiplicity, that is shown
on Fig.~\ref{mult_distri}. In fact, the multiplicity of IMF's with charge $Z > 2$ is
larger in AMD. However, if one selects only sizeable fragments, with charge $Z > 6$, 
closer multiplicities are obtained in the two models 
(dashed lines), but still higher (by about 1 unit) in the AMD case.
This difference may be expected from the fact that the total number
   of nucleons contained in these IMF's in AMD is still larger by about
   15 than in SMF (compare dot-dashed lines in Fig. 6).
\begin{figure}
\vskip 1.0cm
\includegraphics[width=8.cm]{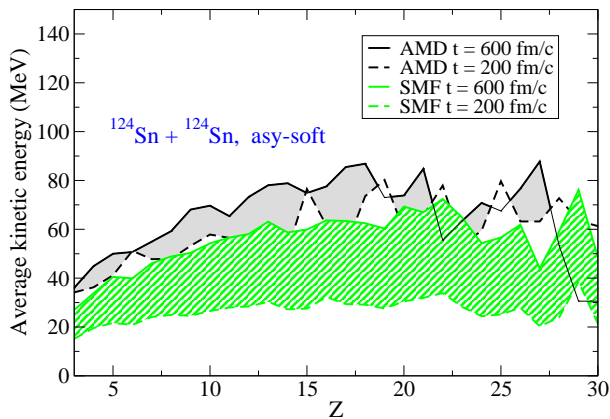}
\caption{IMF average kinetic energy as a function of the charge Z, as obtained
in the two models at $t = 200$ fm/$c$ (dashed lines) and at  $t = 600$ fm/$c$ (full
lines), for the same reaction of  Fig.~\ref{char_distri}.
The solid and hatched grey areas mark the change
during this time interval for AMD and SMF, respectively.
}
\label{ene_kin}
\end{figure}

As discussed above, the average kinetic energy of nucleons emitted at the pre-equilibrium 
stage of the reaction is nearly the same in the two codes. However, the larger amount 
of nucleons emitted in SMF causes 
a considerable reduction of the energy available for the remaining fragmenting system. 
As a consequence, 
the energy (kinetic + internal) stored into IMF's results to be larger in
AMD. More precisely, the fragment intrinsic excitation energy is close
in the two models and amount to 
$\epsilon/A \approx 3 \pm 1$ MeV (corresponding to temperatures
$T\approx 5$ MeV), while  
according also to the different compression-expansion dynamics (see Section
III.A), the fragments have higher collective velocities in AMD. 

The fragment average kinetic energy is represented as a function of the fragment charge 
in Fig.~\ref{ene_kin}. The dashed lines correspond to the time instant
$t = 200$ fm/$c$, 
while full lines are for fragments at  $t = 600$ fm/$c$, where they have been
accelerated by the Coulomb repulsion. One can notice that in the AMD case, these Coulomb
acceleration effects are not so pronounced, since fragments
are rather distant from each other already at $t = 200$ fm/$c$ (see  Fig.\ \ref{contour_BGBD}).
The shape of the average kinetic energy, with an almost linear increase up to charges around $Z=15$, 
denotes the presence of a collective flow velocity, as also experimentally observed in
similar reactions \cite{Indra_nat}. Fragments with larger size, that probably emerge from the coalescence
of two or more smaller objects and are more likely located closer to the system center of
mass, are slowed down.  
The difference between the final average kinetic energy per nucleon  in the two models amounts to
$20\%$.

\subsection{Isospin effects}

\begin{figure}
\vskip 1.0cm
\includegraphics[width=8.cm]{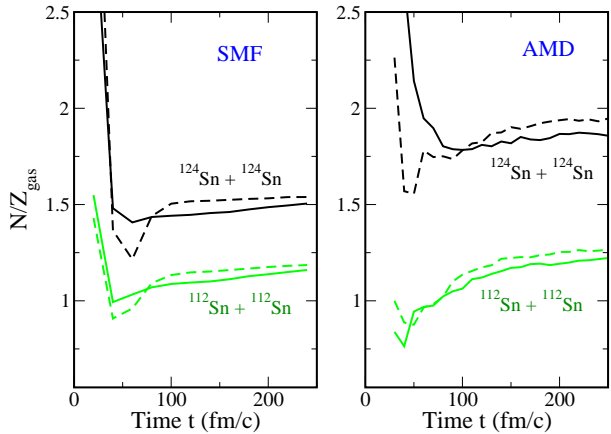}
\caption{Time evolution of the $N/Z$ content of the pre-equilibrium
  emission, as obtained in the two models (left and right panels), 
for the $^{112}\mathrm{Sn} +
  {}^{112}\mathrm{Sn}$ and $^{124}\mathrm{Sn} + {}^{124}\mathrm{Sn}$
  reactions. Full lines: asy-stiff interaction. Dashed lines: asy-soft.}
\label{nzgas}
\end{figure}

It is very interesting to investigate also the isotopic content of
pre-equilibrium and fragment emission, in connection with the asy-EOS
employed and the details of the dynamical models under investigation.
Differences between the two models are seen not only for the abundance
of the pre-equilibrium emission, but also for its isotopic content.
This is illustrated in Fig.~\ref{nzgas}, that shows the time evolution
of the $(N/Z)_{\text{gas}}$ ratio of emitted particles with $A\leq 4$,
in both models, for the two iso-EOS's and the two reactions
considered.  The different calculations present some common features.
In fact, we observe in all cases that the two curves of different
asy-EOS's cross at times $t$ around 70-100 fm/c, which is more evident
for the neutron-rich system.  According to the compression-expansion
dynamics followed by the composite nuclear system, the crossing is
connected to the fact that the system gradually evolves from a compact
shape (compression, high density) where the symmetry energy is larger
and the system likes to emit more neutrons in the asy-stiff case, to a
dilute configuration (expansion, low density) where the  symmetry
energy is larger in the asy-soft case.

\begin{figure}
\vskip 1.0cm
\includegraphics[width=8.cm]{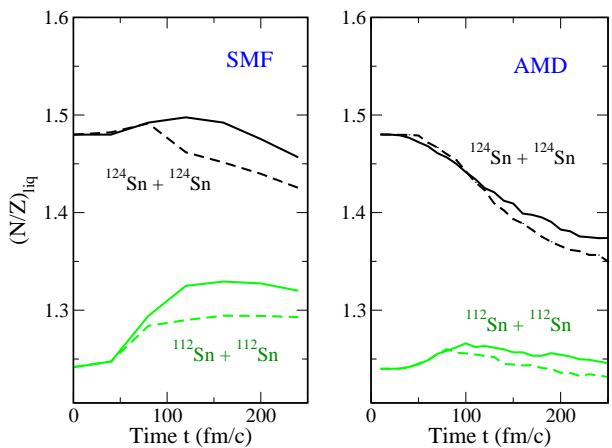}
\caption{Time evolution of the isotopic content of the ``liquid''
  phase, as obtained in the two models (left and right panels), 
for the $^{112}\mathrm{Sn} +
  {}^{112}\mathrm{Sn}$ and $^{124}\mathrm{Sn} + {}^{124}\mathrm{Sn}$
  reactions. Full lines: asy-stiff interaction. Dashed lines: asy-soft.}
\label{z_a_liq}
\end{figure}

The results of Fig.~\ref{nzgas} can be connected to the differences
observed along the compression-expansion dynamics in the two models.
In fact, as discussed in Section III.A, within the SMF trajectory the
composite nuclear system may enter lower density regions as a nearly
homogeneous source and for a longer time, compared to the path
followed in AMD.  Hence the moderate $(N/Z)_{\text{gas}}$ of the
pre-equilibrium emission observed in SMF can be explained by the
rather small value of the symmetry energy at lower density.  This
explanation is also consistent with the more abundant emission
obtained in SMF calculations, since particles are less bound at low
density.  On the other hand, the $(N/Z)_{\text{gas}}$ ratio in the AMD
case is much larger, especially for the neutron-rich
$^{124}\mathrm{Sn}$ reaction.  Due to the efficient clustering effects
in AMD, protons are trapped in fragments that appear at early times,
optimizing the symmetry energy of the whole system.

The same isospin effects can be also observed by looking at the
isotopic content of fragments.  Figure \ref{z_a_liq} shows the
$(N/Z)_{\text{liq}}$ content of the ``liquid'' phase (associated with
the composite nuclear source at early times and with IMF's at later
times), as obtained in the two models and with the two
parametrizations of the symmetry energy.  As expected already from the
results concerning nucleon and light cluster emission, the ``liquid''
is more neutron-rich in SMF.  One can notice that, for the
neutron-poor system in Fig.~\ref{z_a_liq}, the $(N/Z)_{\text{liq}}$
ratio starts to increase at early times, which is due to the repulsive
effect of the Coulomb interaction and is related to the observation in
Fig.~\ref{nzgas} that $(N/Z)_{\text{gas}}$ is smaller than the $N/Z$
ratio of the total system.  This increasing trend of
$(N/Z)_{\text{liq}}$ is more pronounced in the SMF 
model, due to the more abundant emission in this case.
For the
neutron-rich system, the initial increase of $(N/Z)_{\text{liq}}$ is
weak and observed only in the SMF case.  
  The
$(N/Z)_{\text{liq}}$ ratio turns to decrease approximately when
fragments appear                           and neutrons are emitted,
                             lowering the fragment symmetry energy
(isospin distillation \cite{BaranPR410}).  
This effect is more pronounced in the soft case and happens at earlier
times in AMD.
Moreover, isospin distillation is stronger in the neutron-rich system.
  
At the end, one can see from Figs.~\ref{nzgas} and \ref{z_a_liq} that
the difference between the predictions of the two models is larger
than the difference between the results associated with the two
iso-parametrizations.
This observation should be taken as a warning
that the symmetry energy cannot be extracted 
from merely the $N/Z$ content of the
   reaction products in a model independent way.

\section{Conclusions}

In this paper we have undertaken a quantitative comparison of the features observed in
heavy-ion fragmentation reactions at Fermi energies, as predicted by two transport 
models: SMF and AMD. 
As far as observables of experimental interest are concerned,
one significant  discrepancy  between the two models is connected to the amount of
pre-equilibrium emission, i.e.\ the energetic particles that leave the system at the early
stage of the reaction.
In AMD clustering effects appear to be more relevant, 
reducing the amount of free nucleons emitted, compared to SMF, in favour of  
a richer production of primary light IMF's. 
The yield of sizeable primary IMF's ($Z>6$) is close in the two models. However
in SMF fragments with charge around 10 are slightly more abundant, while in AMD
the tail at larger $Z$ (around 20) is more pronounced. 
The shape of the SMF charge distribution is closer to the expectations of spinodal decomposition
\cite{rep}.  However, it should be noticed that
these differences can be  smoothened by secondary decay effects \cite{frag}.  In fact, both models are 
able to fit (final) experimental charge distributions (of IMF's) reasonably well \cite{John,Onoj}.   
In SMF fragment kinetic energies are smaller, compared to AMD, by about 20$\%$. 
These observations corroborate the scenario of a faster fragmentation process in
AMD, while in SMF the system spends a longer time as a nearly homogeneous source 
at low density, emitting a larger amount of nucleons prior to fragment formation.

 
Another interesting feature is that 
the $N/Z$ ratio of the pre-equilibrium emission and of primary IMF's is different in the two
models. For a fixed parametrization of the symmetry potential,   
the isotopic content of the emitted particles is systematically lower in SMF, where
the emission is more abundant. 
As a consequence, IMF's are more neutron-rich in SMF than in AMD. 
The latter observation leads to the conclusion that 
isotopic properties 
are largely affected by the reaction dynamics. 
This can be expected just from the fact that the
symmetry energy is density dependent, so
isospin observables should keep the fingerprints of the density regions spanned in the collision.
However,  the impact of the reaction path on these observables may be more intricate.
From this point of view, isospin properties can be also considered 
as a good tracer of the reaction mechanism  
and may contribute to probe  the corresponding fragmentation path.
Hence, the discrepancy between the two models may be ascribed essentially
to the different compression-expansion dynamics and clusterization effects that,
as shown by our results, affect isoscalar as well as isovector properties of the reaction
products. 

Therefore, the simultaneous analysis of several experimental observables in nuclear
reactions at Fermi energies should help to shed light on the underlying fragmentation
mechanism and the corresponding role of mean-field and many-body correlations.    
In other words, a check of the global reaction dynamics could be a way to test
the validity of the approximations employed in the dynamical models devised to
deal with the complex many-body problem. 
One may look at suitable isoscalar observables
such as fragment and particle yields and 
energy spectra, as well as at the isotopic content of the reaction products.
As shown in Fig.~6, the sharing between the yields of light IMF's and
light particles ($A\leq 4$) is rather different in the two models, the latter
being more abundant and less neutron-rich in SMF.
In particular, 
the study of pre-equilibrium emission, 
which is not influenced  by secondary decay effects,
could really help to 
probe the reaction dynamics, since the difference between AMD and SMF predictions is
rather large, see the results of Fig.s ~6,7,12. 
Only when the model reliability has been established, one may undertake
a deeper investigation of isospin observables, from which more detailed information on
the symmetry energy and its density dependence can be accessed. 
Indeed, as shown in Fig.s~12-13, the
differences between predictions corresponding to two commonly employed symmetry energy 
parametrizations are smaller than the differences associated with the two models considered
here.    



Finally, it would be appealing to extend our comparison of multi-fragmentation
reactions to other approaches introduced to follow the many-body dynamics 
\cite{md,Pap,ImQMD}.  
In such a context, new developments of stochastic
mean field models, in the direction of introducing fluctuations in full
phase space \cite{BL_new} and enhancing the role of correlations, are also of interest.

\acknowledgments 
We thank M.Di Toro for useful comments and discussions. 
We also would like to acknowlegde the European Center for Theoretical Studies in Nuclear Physics and
Related Areas (ECT*, Trento, Italy) for inspiring discussions during workshops on
numerical simulations of heavy ion reactions. 
This work is supported by Grant-in-Aid for Scientific
Research (KAKENHI) 21540253 from Japan Society for the Promotion of
Science, and partly by High Energy Accelerator Research Organization
(KEK) as a supercomputer project.

\end{document}